\documentclass[pra,a4,pacs]{revtex4}

\usepackage{amsfonts}
\usepackage{amsmath}
\usepackage{amssymb}
\usepackage{bm}
\usepackage[dvips]{color}

\usepackage{graphicx}
\usepackage{subfigure}

\newcommand{\hide}[1]{}

\begin{document}


\title{The Dynamics of an Electric Dipole Moment in a Stochastic
Electric Field}

\author{Y. B. Band}

\affiliation{Department of Chemistry, Department of Physics and
Department of Electro-Optics, and the Ilse Katz Center for Nano-Science,
Ben-Gurion University, Beer-Sheva 84105, Israel}


\begin{abstract}
The mean-field dynamics of an electric dipole moment in a
deterministic and a fluctuating electric field is solved to obtain the
average over fluctuations of the dipole moment and the angular
momentum as a function of time for a Gaussian white noise stochastic
electric field.  The components of the average electric dipole moment
and the average angular momentum along the deterministic electric
field direction do not decay to zero, despite fluctuations in all
three components of the electric field.  This is in contrast to the
decay of the average over fluctuations of a magnetic moment in a
stochastic magnetic field with Gaussian white noise in all three
components.  The components of the average electric dipole moment and
the average angular momentum perpendicular to the deterministic
electric field direction oscillate with time but decay to zero, and
their variance grows with time.
\end{abstract}


\maketitle

\section{Introduction}  \label{Sec:Intro}

We consider the decoherence of an electric dipole moment ${\bf d}$ in
an external electric field and in contact with an environment (a bath)
that interacts with it.  Examples of such systems include
heterogeneous diatomic molecules, such as RbCs \cite{Arnaiz_12} and OH
\cite{Stuhl}, polyatomic molecules with a permanent electric dipole
moment (i.e., a molecule, which, if fixed in space so that it cannot
rotate, has a permanent electric dipole moment, even when no external
electric field is present), or a mesoscopic or macroscopic system,
such as a colloidal particle having a dipole moment \cite{Scherer_04}.
The interaction of such systems with an environment can be represented
by evolving the system in an effective electric field, ${\bf
E}^{({\mathrm{eff}})} = {\bf E}_0 + {\bf E}_B(t)$, where ${\bf E}_0$
is the deterministic electric field (which could be time-dependent),
and ${\bf E}_B(t)$ is the electric field which models the influence of
the environment (the bath) on the dipole moment.  The field ${\bf
E}_B(t)$ can be represented by a vector stochastic process
${\boldsymbol \varepsilon}(t)$, where the nature of the environment
determines the type of stochastic process.  Averaging over
fluctuations corresponds to tracing out the environmental degrees of
freedom.  This yields a reduced nonunitary dynamics wherein the
averaged spin decoheres in time.  This approach was recently used to
treat decoherence of spin systems caused by an environment
\cite{STB_2013}.  A prototype model for fluctuations is Gaussian white
noise \cite{vanKampenBook, Kloeden, STB_2013}, wherein the random
process has vanishing correlation time.  We explicitly consider this
prototype noise, although it is simple to use the methods applied here
to treat other kinds of noise, e.g., Gaussian colored noise or
telegraph noise.

It might appear at first sight that the problem of an electric dipole
moment in an electric field having a stochastic contribution is
similar to that of a magnetic dipole moment in a magnetic field having
a stochastic contribution \cite{STB_2013}.  The Stark Hamiltonian for
an electric dipole moment in an electric field is $H_S = -{\bf d}
\cdot {\bf E}$, and the torque it experiences is ${\boldsymbol \tau}_S
= {\bf d} \times {\bf E}$.  This parallels the Zeeman Hamiltonian for
a magnetic dipole moment in a magnetic field, $H_Z = -{\boldsymbol
\mu} \cdot {\bf B}$, and the torque, ${\boldsymbol \tau}_Z =
{\boldsymbol \mu} \times {\bf B}$.  The similarities are striking!
However, there is an important difference.  The electric dipole moment
of a molecule is locked along a molecule-fixed direction (the diatomic
axis in the case of a heterogeneous diatomic molecule), and its
evolution in an electric field is coupled to the rotational motion of
the molecule.  For example, consider the case of a heterogeneous
diatomic molecule of ${}^1 \Sigma$ electronic state symmetry, where
the angular momentum of the molecule is perpendicular to the diatomic
molecule axis, whereas the electric dipole moment is along the
diatomic molecule axis.  In contradistinction, the magnetic moment of
a particle with a magnetic moment is proportional to the angular
momentum of the particle.  The case of an electric dipole moment in an
electric field is more analogous to the case of a magnetic needle in
the presence of a magnetic field \cite{Budker}; the magnetic moment of
the needle is locked by the lattice crystal structure of the needle
along the needle axis.

The present paper considers only {\em one} particle with an electric
dipole moment in a stochastic electric field.  A significant
literature exists on the dynamics of a large collection of particles
with electric dipole moments, as in ferroelectric liquids.
Ferroelectric liquids are analogous to ferromagnetic fluids, also a
well studied topic, wherein the magnetic moments of the individual
particles in the fluid can coherently lock-up, thereby resulting in a
macroscopic magnetic moment \cite{Odenbach}.  The treatment of such
systems in stochastic fields are complicated by interparticle
interactions, making them inherently many-body problems.  The goal of
this work is to develop methods to describe and analyze the dynamics
and decoherence of a single electric dipole moment in a stochastic
field.  Understanding the implications of this work to more
complicated many-body problems would require much further study.

The outline of this paper is as follows.  In Sec.~\ref{Sec:Cl_sol} we
consider the classical dynamics of an electric dipole moment in the
presence of a deterministic and stochastic electric field, and in
Sec.~\ref{SubSec:Stochastic_field} we discuss the dynamics in a
stochastic field.  In Sec.~\ref{Sec:Quantum_Treatment} we develop the
quantum equations of motion of an electric dipole moment in an
electric field, in Sec.~\ref{SubSec:HEM} we present the Heisenberg
equations of motion for the angular momentum and direction of the
dipole, and in Sec.~\ref{SubSec:MFD} we discuss the mean-field
dynamics, which are equivalent to the classical dynamics.  We present
the calculated results in Sec.~\ref{Sec:Calc_results}, and a summary
and conclusion, along with some comments on how to generalize the
treatment beyond the external noise assumption \cite{vanKampenBook}
wherein no back-action of the system on the environment is present is
contained in Sec.~\ref{Sec:Summary}.

\section{Classical Dynamics}  \label{Sec:Cl_sol}

\begin{figure}
\centering
\includegraphics[width=0.4\textwidth]{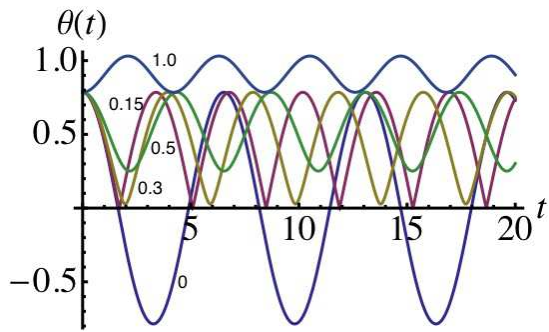}
\caption{(Color online) $\theta(t)$ versus time for several values of
the ratio of parameters $Ed/I$ and $\omega^2$ equal to 0, 0.15, 0.3,
0.5, and 1.}
\label{Fig_Electric_dipole_theta_vs_t}
\end{figure}

Let us begin by considering the classical dynamics of systems having
an electric dipole moment in the presence of an electric field.  For
the moment, let us take the electric field in the direction of the
space-fixed $z$-axis.  The dipole moment ${\bf d}$ in the electric
field ${\bf E}$ experiences a torque, ${\boldsymbol \tau}_S = {\bf d}
\times {\bf E} = E d \, \sin \theta \, \hat {\bm \tau}$, where
$\theta$ is the angle between the electric field and the dipole moment
($\theta$ is the polar angle of the dipole moment).  $\hat {\bm \tau}$
is perpendicular to the $z$-axis, so the $z$ component of angular
momentum, $L_z$, is conserved.  If the system has a moment of inertia
$I$, its angular momentum is ${\bf L} = I {\bm \Omega}$, where ${\bm
\Omega} = \dot \theta \, {\hat {\bm \tau}} + \sin^2 \theta \, \dot
\phi \, {\hat {\bf z}}$ is the angular frequency vector.  We can
denote the conserved $z$-component of the angular momentum as $I
\omega$.  The kinetic energy of the system is given by $T=\tfrac{1}{2}
I (\dot {\theta}^2 + \sin^2 \theta \, \dot {\phi}^2)$, and the Stark
potential energy is $U = - {\bf d} \cdot {\bf E} = - E d \, \cos
\theta$; hence the Lagrangian is
\begin{equation} \label{Lagrangian}
    {\cal L}(\theta, \phi, \dot{\theta}, \dot {\phi}) = T-U =
    \tfrac{1}{2} I (\dot {\theta}^2+\sin^2 \theta \dot {\phi}^2)+ E d
    \, \cos \theta .
\end{equation} 
The Euler-Lagrange equations of motion are
\begin{equation} \label{eq_phi}
    0 = \frac{\partial {\cal L}}{\partial \phi}-\frac{d}{dt}
    \frac{\partial {\cal L}}{\partial \dot{\phi}} = -I \frac{d}{dt}
    (\sin^2 \theta \, \dot{\phi}) \ \ \Rightarrow \ \ \dot{\phi}=
    \frac{\omega}{\sin^2 \theta},
\end{equation} 
\begin{equation} \label{eq_theta}
    0 = \frac{\partial {\cal L}}{\partial \theta}-\frac{d}{dt}
    \frac{\partial {\cal L}}{\partial \dot{\theta}} \ \ \Rightarrow
    \quad \ddot{\theta} +\frac{E d}{I} \sin \theta -\frac{\omega^2}{2}
    \frac{\sin 2 \theta}{\sin^2 \theta} = 0 .
\end{equation} 
The dynamics are relatively simple since the $z$-component of the
angular momentum, $L_z = \frac{\partial {\cal L}}{\partial \dot{\phi}}
= I \sin^2 \theta \, \dot{\phi} \equiv I \omega$, is conserved.  The
second constant of the motion is the total energy ${\cal E}$,
\begin{equation} \label{d}
    {\cal E} = T + U = \tfrac{1}{2} I ( \dot {\theta}^2 +
    \frac{\omega^2}{ \sin^2 \theta} ) - E d \, \cos \theta~.
\end{equation}
Unfortunately, an analytical solution of the differential equation,
$\ddot{\theta} +\frac{E d}{I} \sin \theta -\frac{\omega^2}{2} \sin 2
\theta = 0$, is not known, although for $\omega = 0$, the solution can
be expressed in terms of the Jacobi amplitude for Jacobi elliptic
functions \cite{Abramowitz}, and the solution corresponds to pendular
motion.  For arbitrary $\omega$, the motion is composed of a rotation
around the direction of the electric field ${\bf E}$ with constant
angular velocity $\omega$, and a pendular motion in the plane
containing ${\bf E}$ and ${\bf d}$ (for $\omega \ne 0$, the $\theta$
motion corresponds to distorted pendular motion).
Figure~\ref{Fig_Electric_dipole_theta_vs_t} plots $\theta(t)$ versus
$t$ for several values of the ratio of the parameters $Ed/I$ and
$\omega^2$ in Eq.~(\ref{eq_theta}) for $\theta(t)$.  It is clear from
the figure that a finite $\omega$ keeps $\theta(t)$ away from $\theta
=0$.

\subsection{Stochastic Electric Field} \label{SubSec:Stochastic_field}

Suppose that, in addition to the deterministic electric field, there
is a stochastic electric field contribution, ${\bf E} =
\varepsilon_x(t) \hat{\bf x} + \varepsilon_y(t) \hat{\bf y} + [E +
\varepsilon_z(t)] \hat{\bf z}$, where $\varepsilon_i(t)$, $i = x, y,
z$, are stochastic random variables.  In what follows we explicitly
take Gaussian white noise; see Eqs.~(\ref{white_noise_1}) and
(\ref{white_noise_2}).  Other kinds of noise can occur, e.g., Gaussian
colored noise or telegraph noise, but this paradigm serves to
illuminate the salient features of the dynamics.  Moreover, as long as
the correlation time of the noise is the shortest time scale in the
dynamics, Gaussian white noise is a good approximation for other kinds
of noise.

If only $E_z$ fluctuates, we have to solve the stochastic differential
equation,
\begin{equation}  \label{e_stochastic}
    \ddot{\theta} +\frac{d[E + \varepsilon_z(t)] }{I} \sin \theta
    -\frac{\omega^2}{2} \sin 2 \theta = 0 .
\end{equation}
This equation is linear in $\varepsilon_z(t)$, but nonlinear in
$\theta$.  The stochastic field $\varepsilon_z(t)$ results in a
stochastic variation of the period of the pendular motion in $\theta$.
The addition of stochastic field components $\varepsilon_x(t)$ and
$\varepsilon_y(t)$ result in an additional torque which has a
component along the $z$-axis, i.e., $L_z$ is no longer conserved, and
there is an additional stochastic potential, $V_\perp = -({\bf d}
\cdot {\bf E}_\perp) = -p[\varepsilon_x(t) \sin \theta \cos \phi +
\varepsilon_y(t) \sin \theta \sin \phi]$.  Adding this potential to
the potential $V = - {\bf d} \cdot {\bf E}$, we find $\frac{\partial
{\cal L}}{\partial \phi} = d[-\varepsilon_x(t) \sin \theta \sin \phi +
\varepsilon_y(t) \sin \theta \cos \phi]$, hence
\begin{equation}  \label{phi_stochastic}
    \sin^2 \theta \, \ddot{\phi} + 2 \sin \theta \cos \theta \, \dot
    \theta \, \dot \phi = \frac{d}{I}[-\varepsilon_x(t) \sin \theta
    \sin \phi + \varepsilon_y(t) \sin \theta \cos \phi] .
\end{equation}
The second order differential equations (\ref{e_stochastic}) and
(\ref{phi_stochastic}) can be turned into a set of first order
differential equations.  Defining $\vartheta(t) \equiv \dot
\theta(t)$, Eq.~(\ref{e_stochastic}) becomes,
\begin{equation}  \label{e_stochastic'}
    \frac{d}{dt} \left( \!
    \begin{array}{c}
        \theta(t)  \\
        \vartheta(t)
    \end{array} \! \right) = \left(
    \begin{array}{c}
        \vartheta(t) \\
	- \frac{d[E + \varepsilon_z(t)] }{I} \sin \theta
	+ \frac{\omega^2}{2} \sin 2 \theta \end{array} \right) ,
\end{equation}
and, defining $\varphi(t) \equiv \dot \phi(t)$,
Eq.~(\ref{phi_stochastic}) is transformed into the first order set of
equations,
\begin{equation}  \label{phi_stochastic'}
    \frac{d}{dt} \left( \!
	\begin{array}{c}
	    \phi(t)  \\
	    \varphi(t)
	\end{array} \! \right) =  \left(
    \begin{array}{c}
        \varphi(t) \\
    - 2 \frac{\sin \theta \cos \theta}{\sin^2 \theta} \, \dot \vartheta \,
    \dot \varphi + \frac{d}{I \sin^2 \theta}[-\varepsilon_x(t) \sin \theta
    \sin \phi + \varepsilon_y(t) \sin \theta \cos \phi] \end{array}
    \right) .
\end{equation}
Equation~(\ref{e_stochastic'}) can be solved first for $\theta(t)$ and
$\vartheta(t)$, and these functions can the be substituted into
Eq.~(\ref{phi_stochastic'}), which can then be used to obtain
$\phi(t)$ and $\varphi(t)$.  Alternatively, Eqs.~(\ref{e_stochastic'})
and (\ref{phi_stochastic'}) can be solved simultaneously as a system
of four first-order differential equations.

\section{Quantum Treatment}  \label{Sec:Quantum_Treatment}

The Hamiltonian for the system is given by $H = \frac{ {\hat L}^2}{2
I} - {\bf d} \cdot {\bf E}$ \cite{sym_top}.  In spherical coordinates,
if the electric field is taken to be along the $z$-axis, the
Hamiltonian takes the form,
\begin{equation} \label{Ham_theta_phi}
    H = - \hbar^2 \left( \frac{\partial^2}{\partial \theta^2} + \cot
    \theta \frac{\partial}{\partial \theta}+\frac{1}{\sin^2
    \theta}\frac{\partial^2}{\partial \phi^2} \right) - E d \, \cos
    \theta .
\end{equation} 
Since ${\hat L}_z$ is conserved if the electric field is along the
$z$-axis, the eigenfunctions of the Hamiltonian (\ref{Ham_theta_phi})
can be written as $\psi_{nm}(\theta,\phi) = e^{i m \phi}
f_{nm}(\theta)$, where the functions $f_{nm}(\theta)$ satisfy the
stationary Schr\"odinger equation in one variable,
\begin{equation} \label{Shr_theta_phi}
    \left[- \frac{ \hbar^2 }{2 I} \left( \frac{\partial^2}{\partial
    \theta^2} + \cot \theta \frac{\partial}{\partial
    \theta}+\frac{m^2}{\sin^2 \theta} \right) + E d \, \cos \theta
    \right] f_{nm}(\theta) = {\cal E}_{nm}f_{nm}(\theta) .
\end{equation}
The energy eigenvalues ${\cal E}_{nm}$ have quadratic and higher
contributions in the electric field strength.

If a degeneracy of the energy levels having different angular momentum 
is present, as occurs for molecules with $\Pi$ or higher electronic
symmetry, a Stark energy which is linear in the electric field strength 
can arise.  We shall not consider the dynamics for such cases here.

If, in addition to the constant electric field, a time-dependent field
${\boldsymbol \varepsilon}(t) = \varepsilon_x(t) \hat{\bf x} +
\varepsilon_y(t) \hat{\bf y} + \varepsilon_z(t) \hat{\bf z}$ is
present, the time-dependent Schr\"{o}dinger equation must be used.  A
basis of states could be used to calculate the time-dependent wave
function that is the solution to the time-dependent Schr\"{o}dinger
equation.  The basis could be composed of field-free basis states
$Y_{lm}(\theta,\phi)$, or the eigenstates in the presence of the
constant electric field, $\psi_{nm}(\theta,\phi)$.  Let us now
consider the quantum treatment of the dynamics.  The approach we use
below uses instead the Heisenberg equations of motion, which will be
solved in a mean-field approximation.

\subsection{Heisenberg Equations of Motion} \label{SubSec:HEM}

When no internal angular momentum is present \cite{sym_top}, we take
the Hamiltonian to be
\begin{equation}  \label{Ham_sym_top_E_field}
    H = H_0 + H_S = \frac{{\hat L}^2}{2 I} - \hat{{\bf d}} \cdot {\bf
    E} .
\end{equation}
and we use the notation, $\hat{{\bf d}} = d \, \hat{{\bf n}}$ (see
Ref.~\cite{LL_QM}), where $\hat{{\bf n}}$ is a vector operator of unit
length in the direction of the dipole moment and $d$ is the
magnitude of the dipole moment, which remains constant.  The
Heisenberg equations of motion for the dipole moment operator,
$\dot{\hat{{\bf d}}} = \frac{i}{\hbar} [H, \hat{{\bf d}}]$, can be
written as
\begin{equation} \label{dot_n}
    \dot{\hat{{\bf n}}} = \frac{i}{\hbar} [H, \hat{{\bf n}}] =
    \frac{i}{2 \hbar I}[{\hat L}^2, \hat{{\bf n}}] - \frac{i \,
    d}{\hbar} {\bf E} \cdot [\hat{{\bf n}}, \hat{{\bf n}}] .
\end{equation}
Using the fact that $[\hat{L}_i, \hat{n}_j] = i \hbar \epsilon_{ijk}
\hat{n}_k$, we find that $[{\hat L}^2, \hat{{\bf n}}] = 2 i \hbar [
\hat{{\bf L}} \times \hat{{\bf n}} + i \hbar \, \hat{{\bf n}} ]$.
Since $[\hat{n}_i, \hat{n}_j] = 0$ for all $i$ and $j$, we find,
\begin{equation} \label{dot_n_final}
    \dot{\hat{{\bf n}}} = - \frac{1}{I} [ \hat{{\bf L}} \times
    \hat{{\bf n}} + i \hbar \, \hat{{\bf n}} ] .
\end{equation}
Moreover, the torque on the molecule due to the presence of the
external field is, $\dot{\hat{{\bf L}}} = \frac{i}{\hbar} [H,
\hat{{\bf L}}]$, which reduces to
\begin{equation} \label{dot_L}
    \dot{\hat{{\bf L}}} = - d ({\bf E} \times \hat{{\bf n}}) .
\end{equation}
The nonlinear Heisenberg operator equations of motion,
Eqs.~(\ref{dot_n_final}) and (\ref{dot_L}), must be solved
simultaneously.

\subsection{Mean-Field Dynamics}  \label{SubSec:MFD}

If the initial angular momentum of the molecule is large compared to
$\hbar$, a semiclassical treatment can be a good approximation.
Setting $\hbar = 0$ in Eq.~(\ref{dot_n_final}) allows a semiclassical
solution for the expectation values $\langle {\hat{{\bf n}}}(t)
\rangle$ and $\langle {\hat{{\bf L}}}(t) \rangle$.  The semiclassical
equations,
\begin{equation} \label{dot_n_semi}
    \frac{d}{dt} \langle {\hat{{\bf n}}}\rangle = - \frac{1}{I} \,
    \langle {\hat{{\bf L}}}\rangle \times \langle {\hat{{\bf
    n}}}\rangle ,
\end{equation}
\begin{equation} \label{dot_L_repeat}
    \frac{d}{dt} \langle {\hat{{\bf L}}}\rangle = - d \, {\bf E}
    \times \langle {\hat{{\bf n}}}\rangle .
\end{equation}
are equivalent to the classical solution presented in
Sec.~\ref{Sec:Cl_sol}, but are valid for arbitrary direction of ${\bf
E}$.  These equations correspond to a mean-field theory treatment
obtained by taking the expectation values of Eqs.~(\ref{dot_n_final})
and (\ref{dot_L}), replacing the expectation value of the product
$\hat{{\bf L}} \times \hat{{\bf n}}$ by the product of the expectation
values \cite{Zobay_00, Liu_02, Tikhonenkov_06, Band_07} and taking the
limit as $\hbar \to 0$ on the RHS of (\ref{dot_n_final}).

In what follows, we shall simplify the notation and not explicitly 
write the expectation values around the dynamical variables.

\section{Calculated Results}  \label{Sec:Calc_results}

We now present results for the semiclassical dynamics of a dipole
moment in the presence of an electric field, with and without a
stochastic contribution.

\begin{figure} 
\centering
\includegraphics[width=0.6\textwidth]{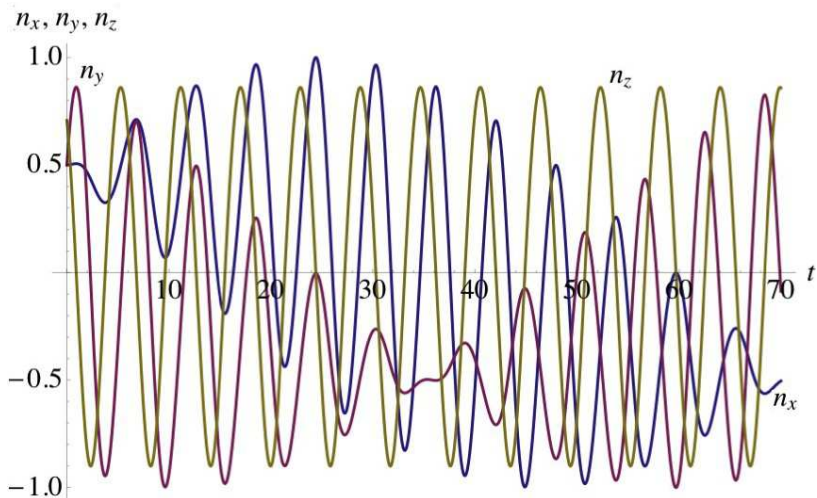}
\caption{(Color online) $n_x(t), n_y(t)$, and $n_z(t)$ versus time for
${\bf n}(0) = (\sin(\pi/4) \cos(\pi/4),\sin(\pi/4)
\sin(\pi/4),\cos(\pi/4))$ and ${\bf L}(0) = (10,0,0)$.  The motion is
almost periodic with period $\approx 140$, but only about half this
region is plotted to minimize congestion.}
\label{Fig_n_vs_t_in_E_field}
\end{figure}

\begin{figure} 
\centering
\includegraphics[width=0.45\textwidth]{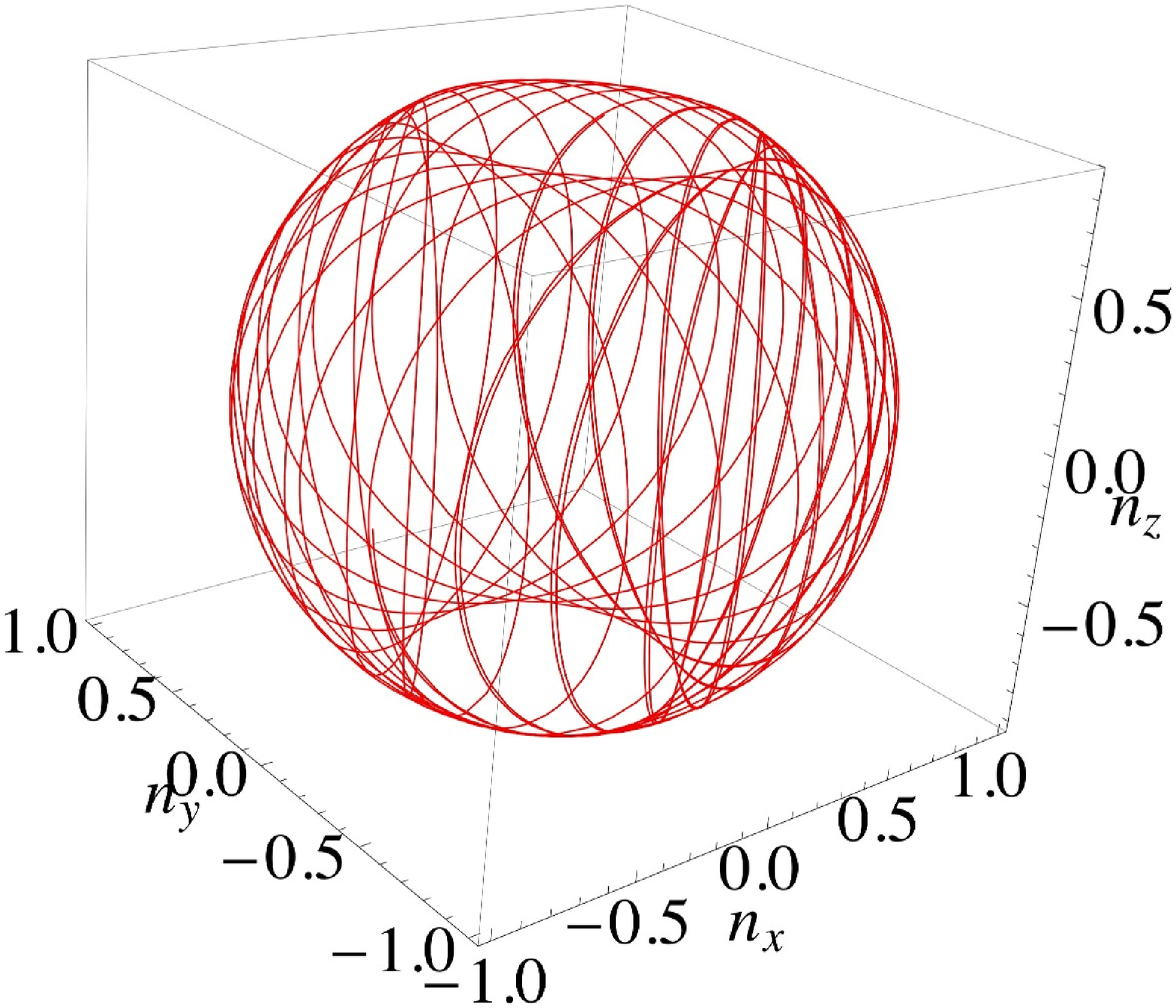}
\caption{(Color online) Parametric plot of $n_x(t), n_y(t)$, and
$n_z(t)$ versus time for ${\bf n}(0) = (\sin(\pi/4)
\cos(\pi/4),\sin(\pi/4) \sin(\pi/4),\cos(\pi/4))$ and ${\bf L}(0) =
(10,0,0)$.  The motion is almost periodic (see the slight differences
in the trajectory upon making two passes) with dimensionless period
$\approx 140$.}
\label{Fig_dipole_in_E_field_vs_t}
\end{figure}

\begin{figure} 
\centering
\includegraphics[width=0.45\textwidth]{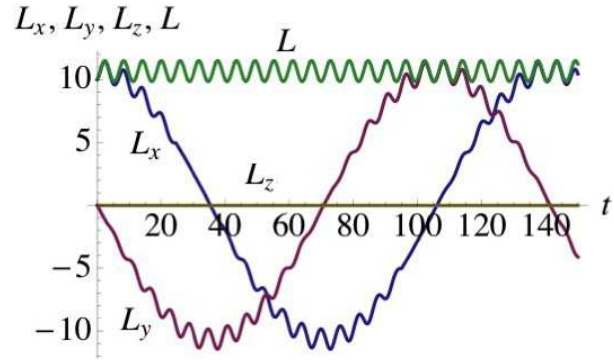}
\caption{(Color online) $L_x(t), L_y(t), L_z(t)$ and $L(t) =
\sqrt{L^2_x(t) + L^2_y(t) + L^2_z(t)}$ versus time for ${\bf n}(0) =
(\sin(\pi/4) \cos(\pi/4),\sin(\pi/4) \sin(\pi/4),\cos(\pi/4))$ and
${\bf L}(0) = (10,0,0)$.  $L_z(t)$ remains identically zero throughout
the dynamics.  The motion is almost periodic with dimensionless period
$\approx 140$.}
\label{Fig_L_vs_t_in_E_field}
\end{figure}

Figures~\ref{Fig_n_vs_t_in_E_field} and
\ref{Fig_dipole_in_E_field_vs_t} show $n_x(t), n_y(t)$, and $n_z(t)$
versus time, and Fig.~\ref{Fig_L_vs_t_in_E_field} shows $L_x(t),
L_y(t)$, and $L_z(t)$ versus time for deterministic dynamics (without
a stochastic contribution) of a dipole moment in an electric field.
The dimensionless parameters used in these calculations are $d = 1$,
$I = 10$, and $(E_x, E_y, E_z) = (0,0,1)$.  The initial conditions are
specified in the figure captions.  $n_z(t)$ undergoes periodic motion,
but the $n_x(t)$ and $n_y(t)$ trajectories are more complicated and
are not truly periodic.  Nevertheless, the motion is almost periodic
with period $\tau = 140$ for this case.  The parametric plot of ${\bf
n}(t)$ in Fig.~\ref{Fig_dipole_in_E_field_vs_t} shows the holes around
the north and south poles.  Figure~\ref{Fig_L_vs_t_in_E_field} shows
that the total angular momentum is not conserved, but $L_z$ remains
zero throughout the dynamics.  The only component of the angular
momentum that is initially nonzero is $L_x$.  The angular momentum
components $L_x(t)$ and $L_y(t)$ undergo a complicated oscillatory
motion as a function of time.

\begin{figure} 
\centering
\includegraphics[width=0.6\textwidth]{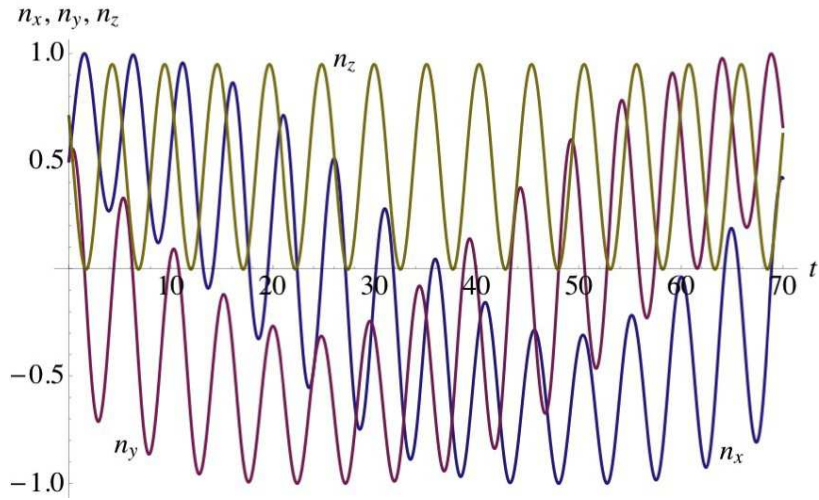}
\caption{(Color online) $n_x(t), n_y(t)$, and $n_z(t)$ versus time for
${\bf n}(0) = (\sin(\pi/4) \cos(\pi/4),\sin(\pi/4)
\sin(\pi/4),\cos(\pi/4))$ and ${\bf L}(0) = (10,0,6)$.  The motion is
almost periodic with dimensionless period $\approx 100$, but only
about 70\% of a period is plotted to minimize congestion.}
\label{Fig_n_vs_t_in_E_field_Lz}
\end{figure}

\begin{figure} 
\centering
\includegraphics[width=0.45\textwidth]{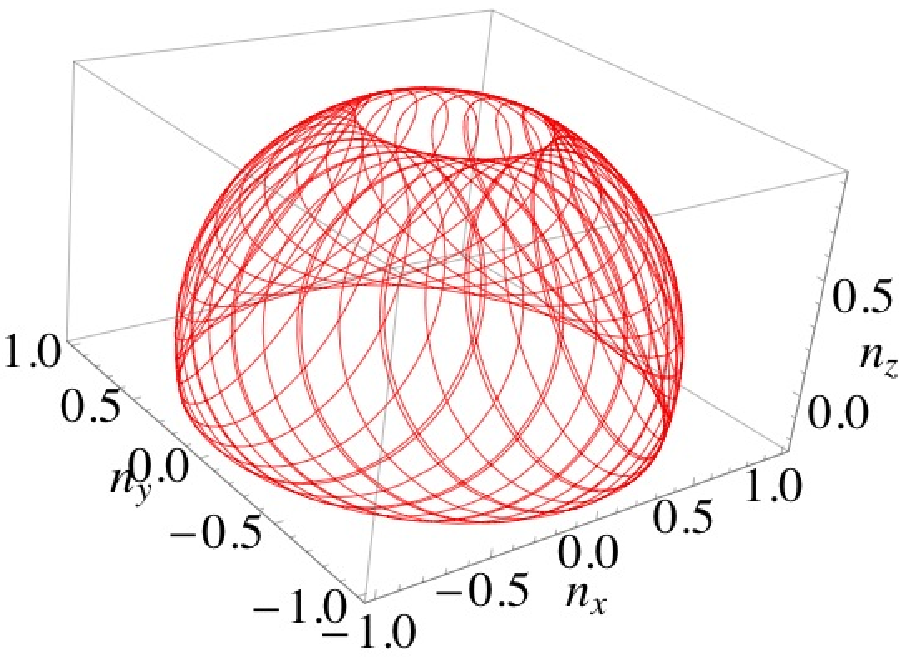}
\caption{(Color online) Parametric plot of $n_x(t), n_y(t)$, and
$n_z(t)$ versus time for ${\bf n}(0) = (\sin(\pi/4)
\cos(\pi/4),\sin(\pi/4) \sin(\pi/4),\cos(\pi/4))$ and ${\bf L}(t) =
(10,0,6)$.  The motion is almost periodic (see the slight differences
in the trajectory upon making two passes) with period $\approx 100$.}
\label{Fig_dipole_in_E_field_vs_t_Lz}
\end{figure}

\begin{figure} 
\centering
\includegraphics[width=0.45\textwidth]{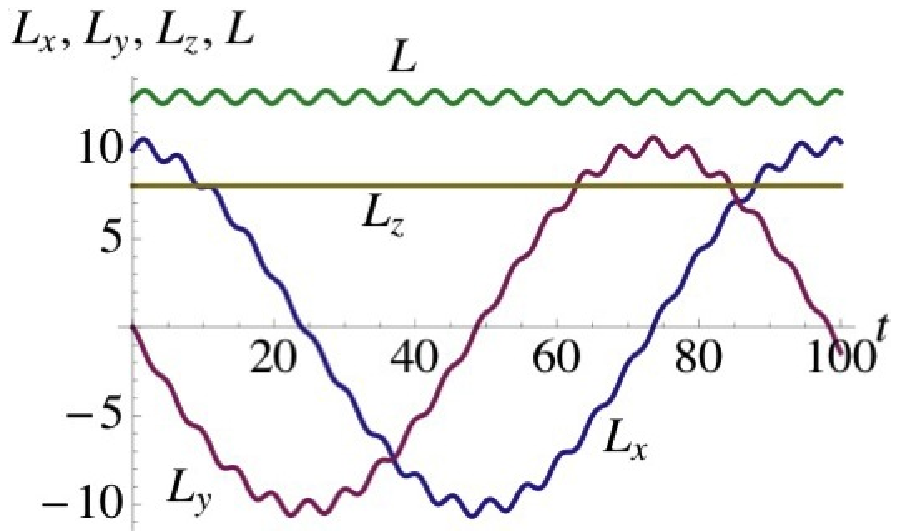}
\caption{(Color online) $L_x(t), L_y(t), L_z(t)$ and $L(t) =
\sqrt{L^2_x(t) + L^2_y(t) + L^2_z(t)}$ versus time for ${\bf n}(0) =
(\sin(\pi/4) \cos(\pi/4),\sin(\pi/4) \sin(\pi/4),\cos(\pi/4))$ and
${\bf L}(0) = (10,0,6)$.  $L_z(t)$ remains equal to $6$ throughout the
dynamics.  The motion is almost periodic with period $\approx 100$.}
\label{Fig_L_vs_t_in_E_field_Lz}
\end{figure}

Figures~\ref{Fig_n_vs_t_in_E_field_Lz} and
\ref{Fig_dipole_in_E_field_vs_t_Lz} show $n_x(t), n_y(t)$, and
$n_z(t)$, and Fig.~\ref{Fig_L_vs_t_in_E_field_Lz} shows $L_x(t),
L_y(t)$, and $L_z(t)$ versus time for deterministic dynamics for the
same conditions as previously, except that now $L_z(0) = 6$, rather
than zero.  Now, the $z$-component of angular momentum, which is
conserved, restricts the values of $n_z(t)$ to be non-negative.  The
motion is again almost periodic, with a dimensionless period of about
$\tau = 100$.

\begin{figure} 
\centering
\centering\subfigure[]{\includegraphics[width=0.45\textwidth]
{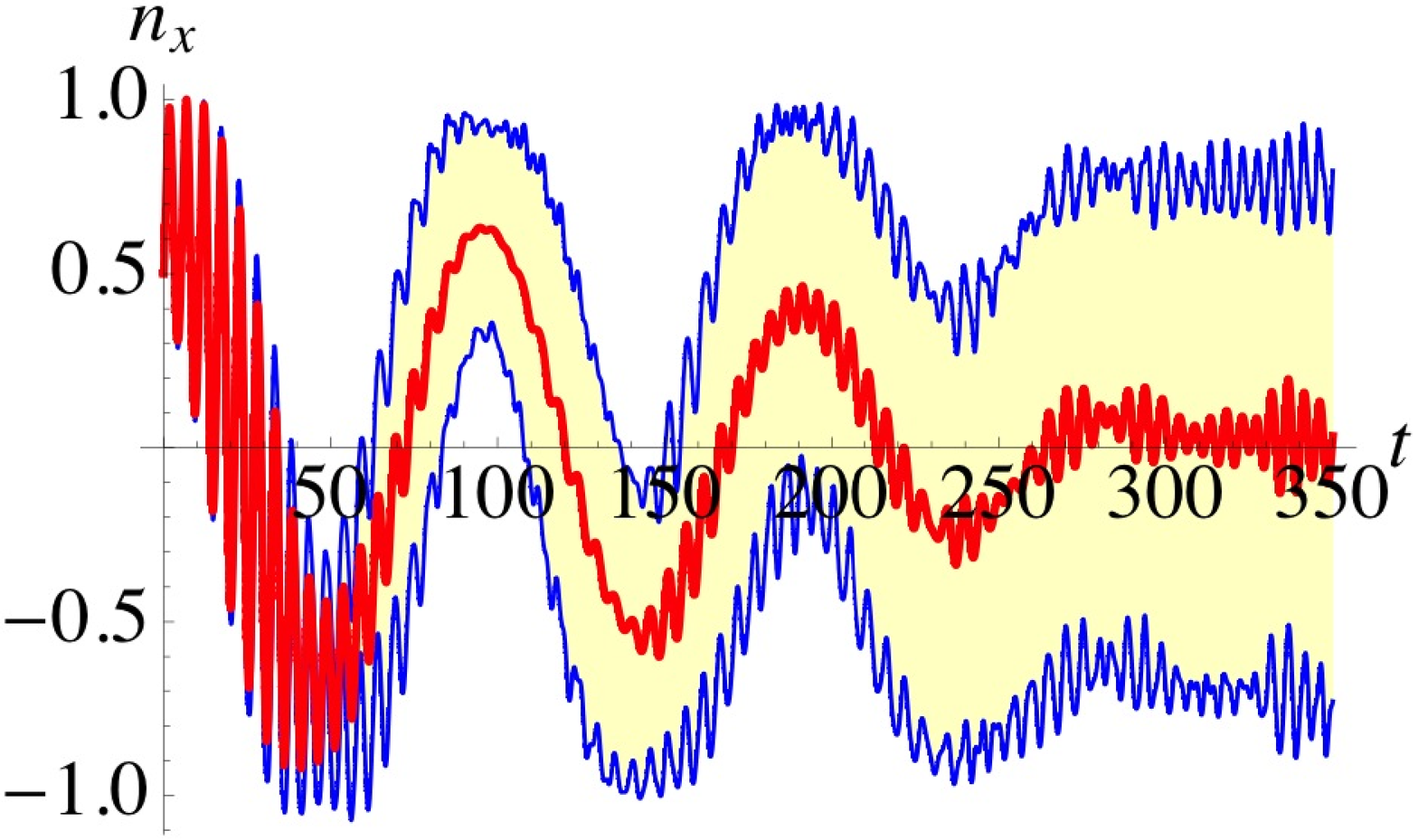}}
\centering\subfigure[]{\includegraphics[width=0.45\textwidth]
{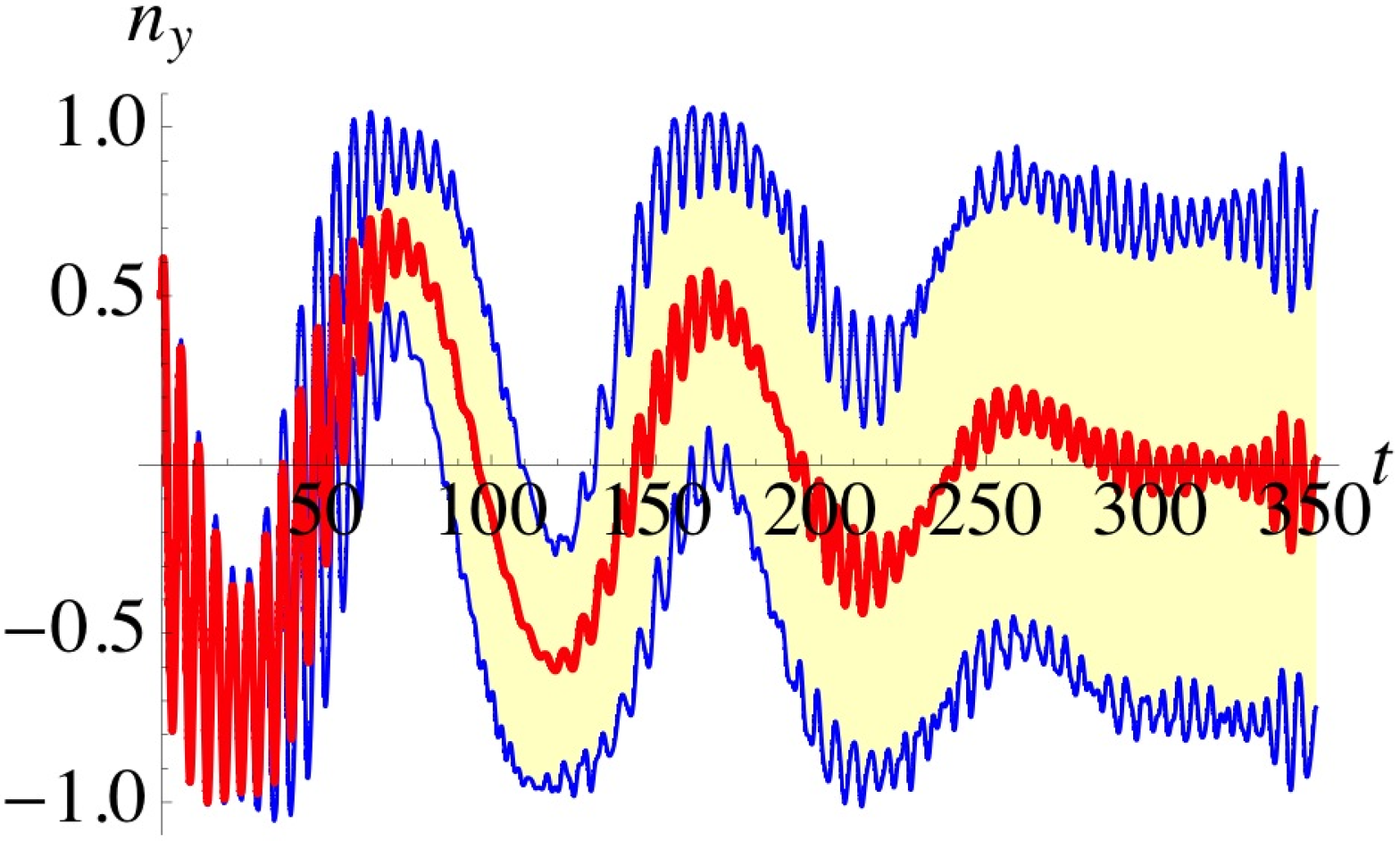}} 
\centering\subfigure[]{\includegraphics[width=0.45\textwidth]
{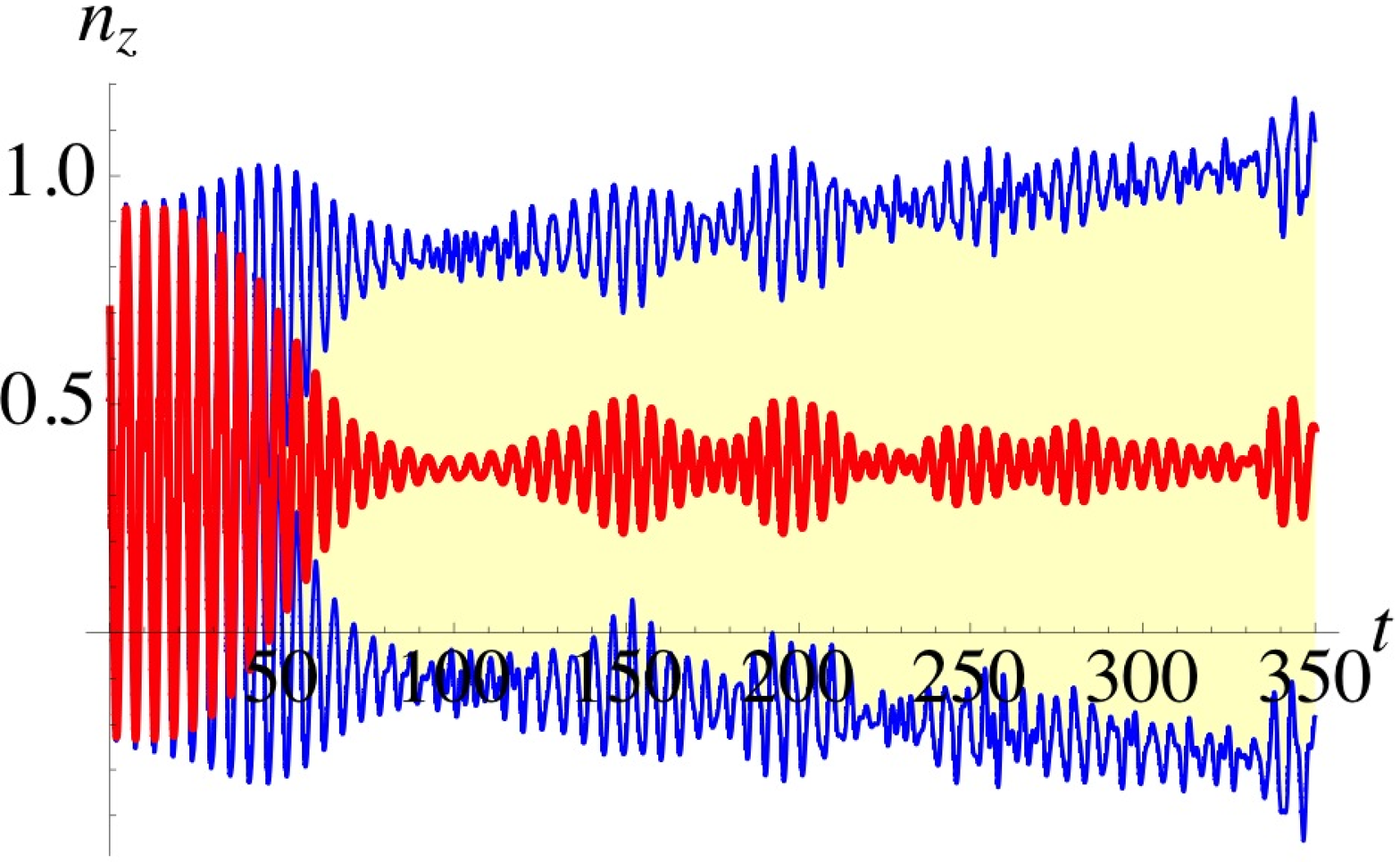}} \caption{(Color online) $n_x(t), n_y(t),
n_z(t)$ versus time obtained for stochastic dynamics with
$\varepsilon_x(t)$ and $\varepsilon_y(t)$ fields taken as Gaussian
white noise.  The initial conditions are ${\bf n}(0) = (\sin(\pi/4)
\cos(\pi/4),\sin(\pi/4) \sin(\pi/4),\cos(\pi/4))$ and ${\bf L}(0) =
(10,0,6)$.  }
\label{Fig_n_vs_t_in_E_field_m_v}
\end{figure}

\begin{figure} 
\centering
\centering\subfigure[]{\includegraphics[width=0.4\textwidth]
{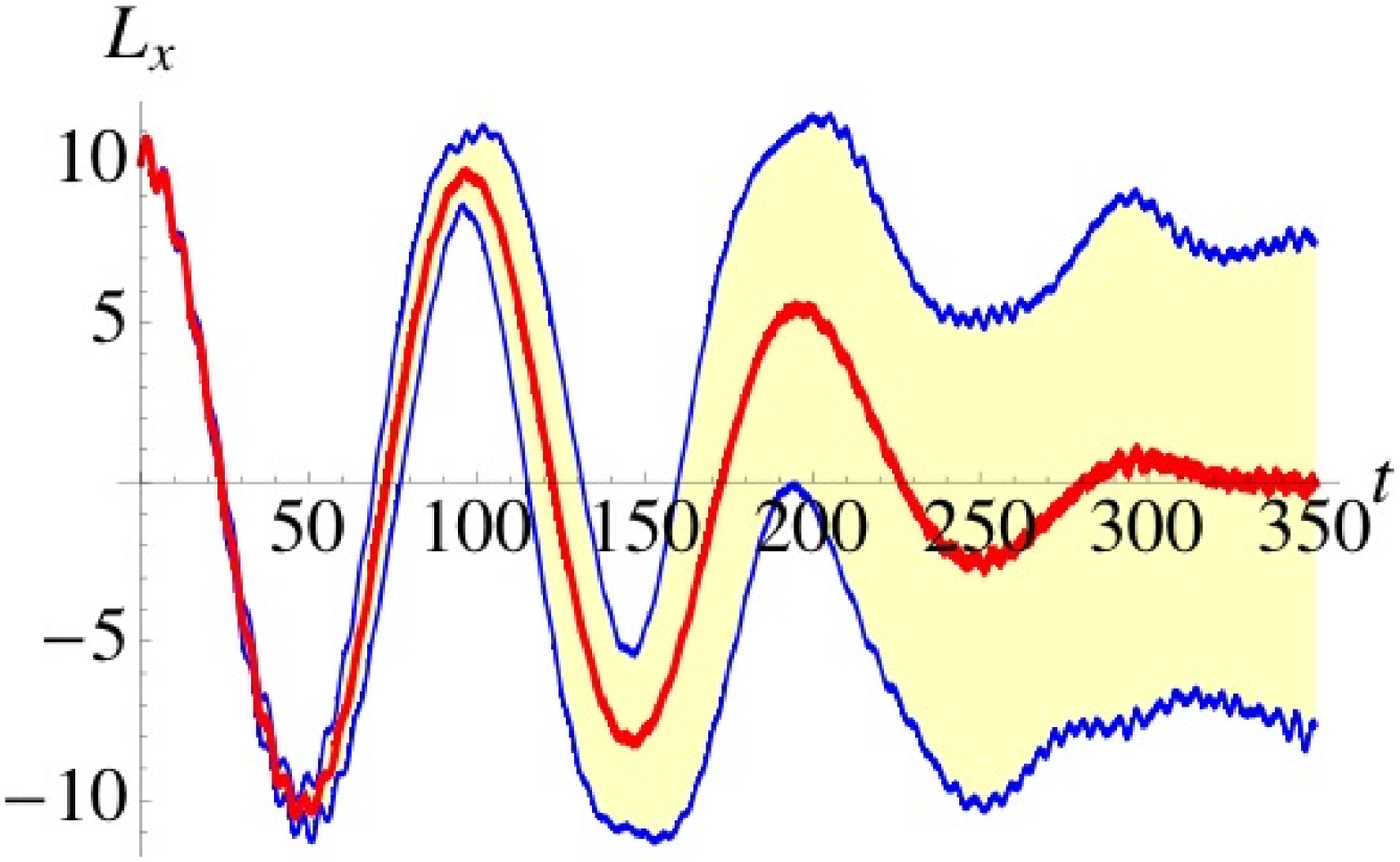}}
\centering\subfigure[]{\includegraphics[width=0.4\textwidth]
{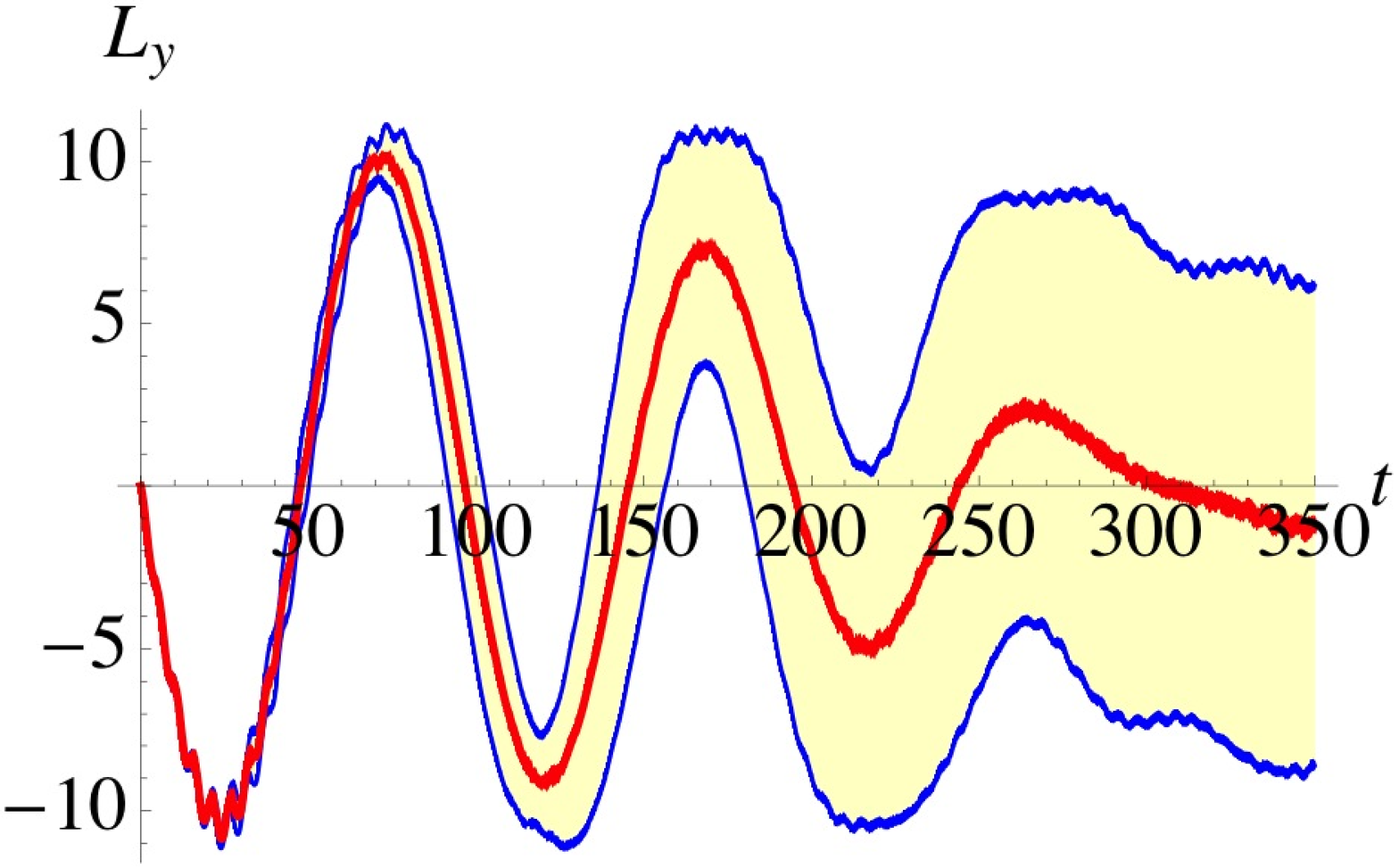}} 
\centering\subfigure[]{\includegraphics[width=0.4\textwidth]
{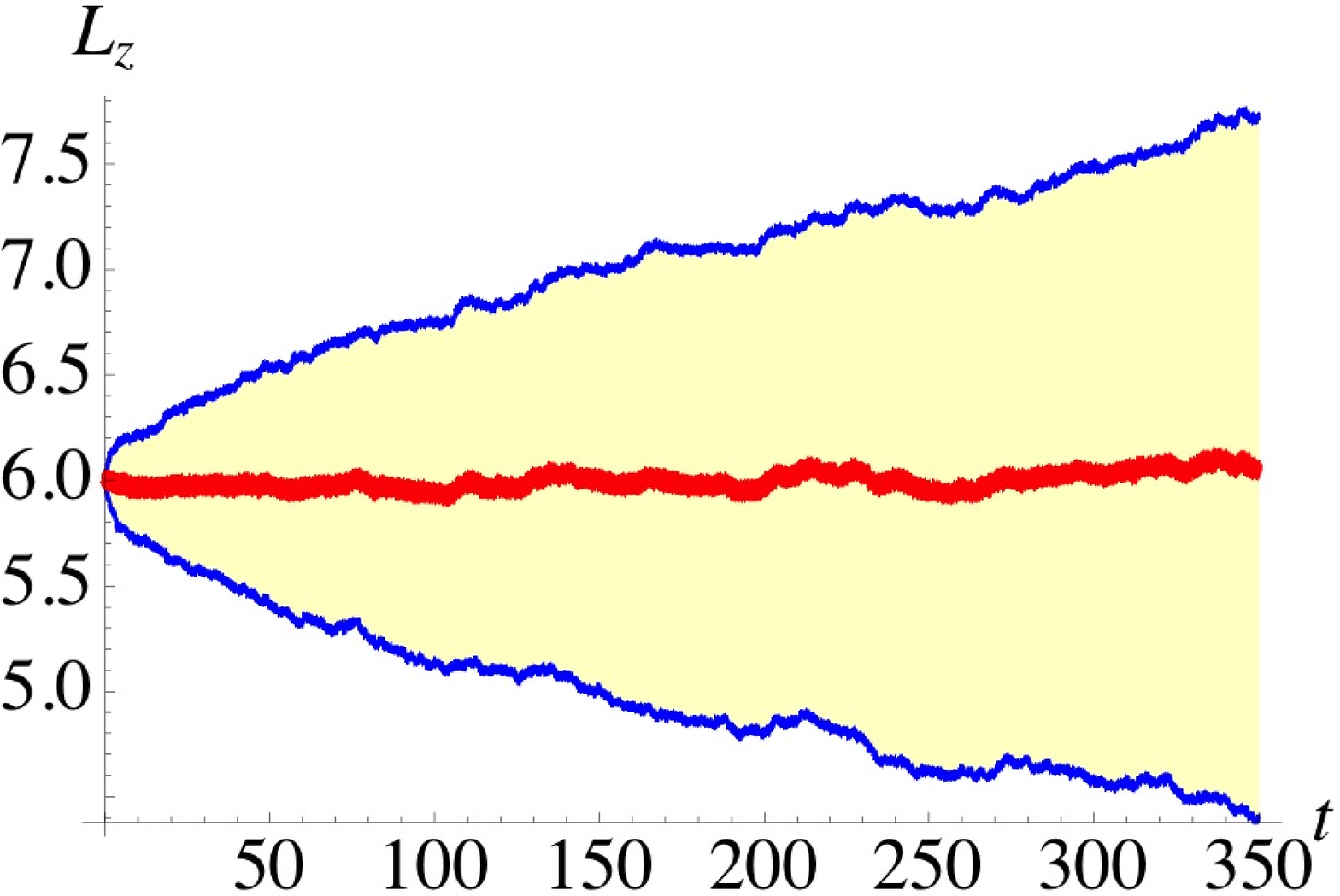}}
\caption{(Color online) $L_x(t), L_y(t), L_z(t)$ versus time obtained
for the stochastic dynamics with $\varepsilon_x(t)$ and
$\varepsilon_y(t)$ fields taken as Gaussian white noise.}
\label{Fig_L_vs_t_in_E_field_m_v}
\end{figure}

We now consider the details of the stochastic electric field.  We take
$\varepsilon_x(t)$ and $\varepsilon_y(t)$ to be stochastic processes
with zero mean and correlation function $\kappa(t-t')$ taken to be a
$\delta$ function,
\begin{equation}  \label{white_noise_1}
    \overline{\varepsilon_i(t)} = 0 ,
\end{equation}
\begin{equation}  \label{white_noise_2}
    \overline{\varepsilon_i(t) \varepsilon_j(t')} = \kappa(t-t') \, 
    \delta_{ij} = \varepsilon_0^2 \, \delta (t-t') \, \delta_{ij} ,
\end{equation}
for $i, j = x, y$, i.e., we consider Gaussian white noise in the
$x$-$y$ plane.  The overline indicates the average over the
fluctuations, and $\delta_{ij}$ is the Kronecker delta function (the
noise in the $x$ and $y$ directions are uncorrelated).  We take the
correlation function $\kappa(t-t')$ to have vanishing correlation
time, $\tau_c = 0$, i.e., Gaussian white noise.  We set the strength
of the fluctuations, $\varepsilon_0$, to be a tenth of the dc electric
field $E_z$ with `volatility' (standard deviation) $\varepsilon_0 =
0.1$, and initially take the fluctuations in the $z$-component to
vanish.
The equations of motion for the stochastic case are written as
\begin{equation} \label{dot_n_semi'}
    d \langle {\hat{{\bf n}}}\rangle = - \frac{1}{I} \,
    \langle {\hat{{\bf L}}}\rangle \times \langle {\hat{{\bf
    n}}}\rangle \, dt,
\end{equation}
\begin{equation} \label{dot_L_repeat'}
    d \langle {\hat{{\bf L}}}\rangle = - d \, ({\bf E} \, dt + d {\bf
    W}) \times \langle {\hat{{\bf n}}}\rangle .
\end{equation}
where ${\bf W}(t)$ is a vector Wiener process.  The white noise,
${\boldsymbol \varepsilon}(t)$ can be written as the time derivative
of the Wiener process, ${\boldsymbol \varepsilon}(t) = d{\bf W}/dt$,
or more formally, the Wiener process is the integral of the white
noise.  The other parameters and initial conditions are taken to be
exactly as in the previous case.  The stochastic field results were
obtained using the Mathematica 9.0 built-in command {\em ItoProcess}
for solving stochastic differential equations, with the stochastic
field ${\bf W}(t)$ taken as a Wiener process.
Figure~\ref{Fig_n_vs_t_in_E_field_m_v} shows $\overline{n_x(t)},
\overline{n_y(t)}$, and $\overline{n_z(t)}$ versus time and
Fig.~\ref{Fig_L_vs_t_in_E_field_m_v} shows $\overline{L_x(t)},
\overline{L_y(t)}$, and $\overline{L_z(t)}$ versus time for the
stochastic dynamics.  In these figures, the mean values and the mean
values plus and minus the standard deviations are shown, and the
region between the plus and minus standard deviations are shaded.  The
standard deviation of $\overline{n_x(t)}, \overline{n_y(t)}$, and
$\overline{n_z(t)}$ become significant for times greater than about
70, whereas the standard deviation of $\overline{L_x(t)},
\overline{L_y(t)}$, and $\overline{L_z(t)}$ become significant only
for times greater than about 150.  The mean values of
$\overline{n_x(t)}$ and $\overline{n_y(t)}$ decay to zero with time,
but $\overline{n_z(t)}$ does not decay to zero (or at least not on the
time scale shown in the figure).  For all $n_i(t)$, $i = x, y, z$, the
standard deviation increases with time, but the increase is slow at
large times.  Moreover, the mean values $\overline{L_x(t)}$ and
$\overline{L_y(t)}$ decay to zero at large time, but
$\overline{L_z(t)}$ hardly decreases on the timescale shown, and the
standard deviation of $L_z(t)$ increases linearly with time at large
times.  We conclude that, despite the fluctuations,
$\overline{n_z(t)}$ and $\overline{L_z(t)}$ do not decay to zero as do
the other components of $\overline{{\bf n}(t)}$ and $\overline{{\bf
L}(t)}$.

\begin{figure} 
\centering
\centering\subfigure[]{\includegraphics[width=0.4\textwidth]
{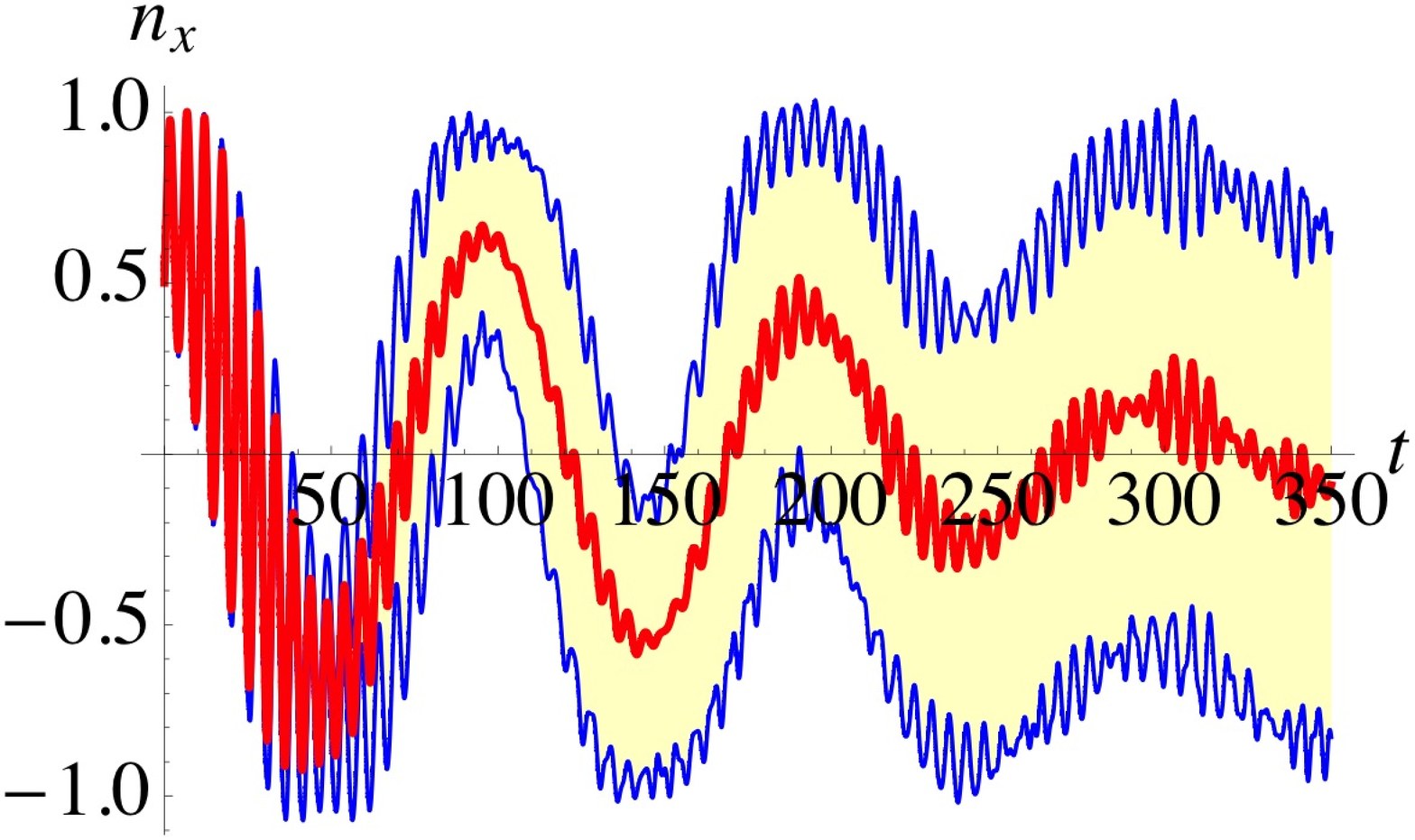}}
\centering\subfigure[]{\includegraphics[width=0.4\textwidth]
{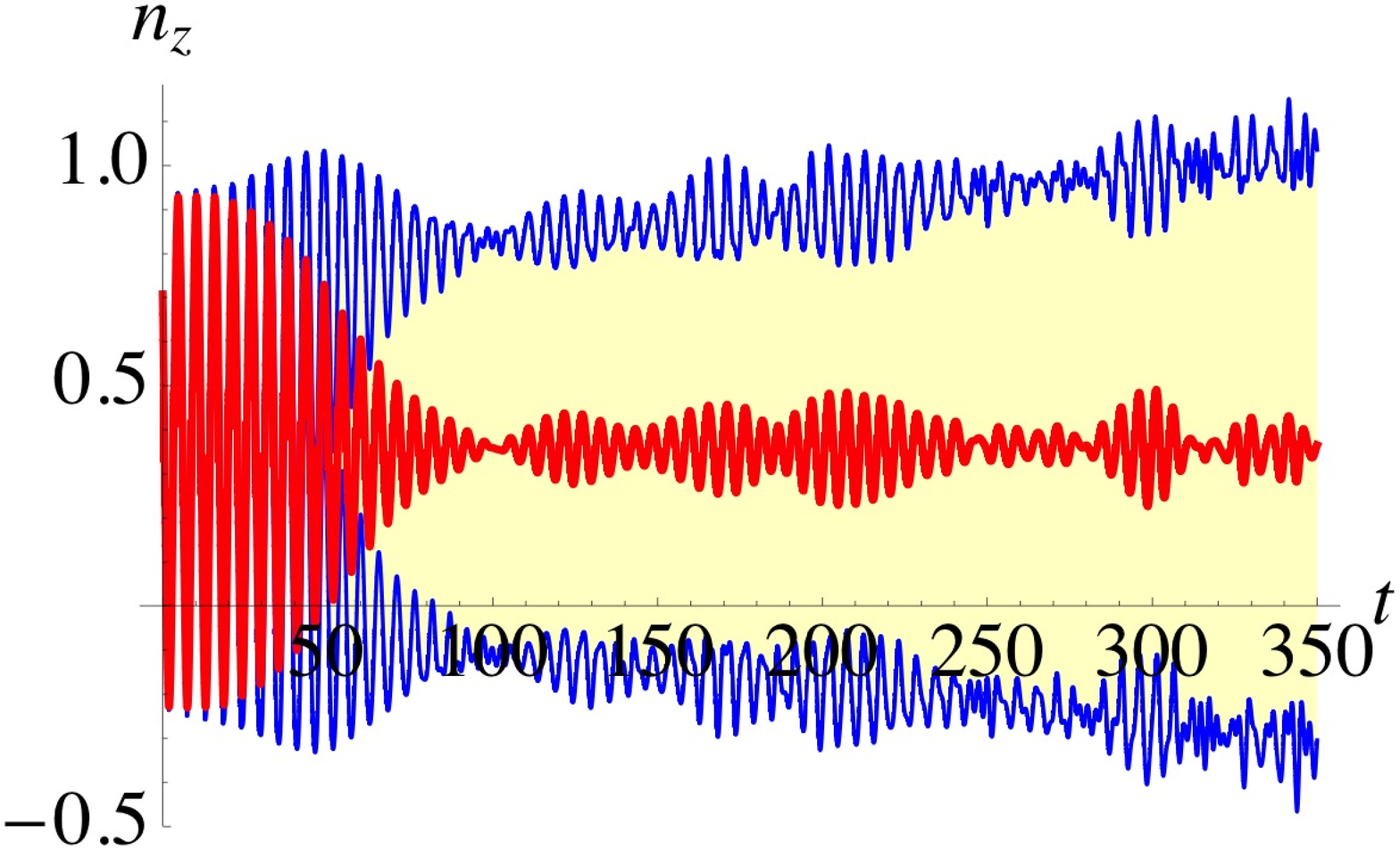}} 
\centering\subfigure[]{\includegraphics[width=0.4\textwidth]
{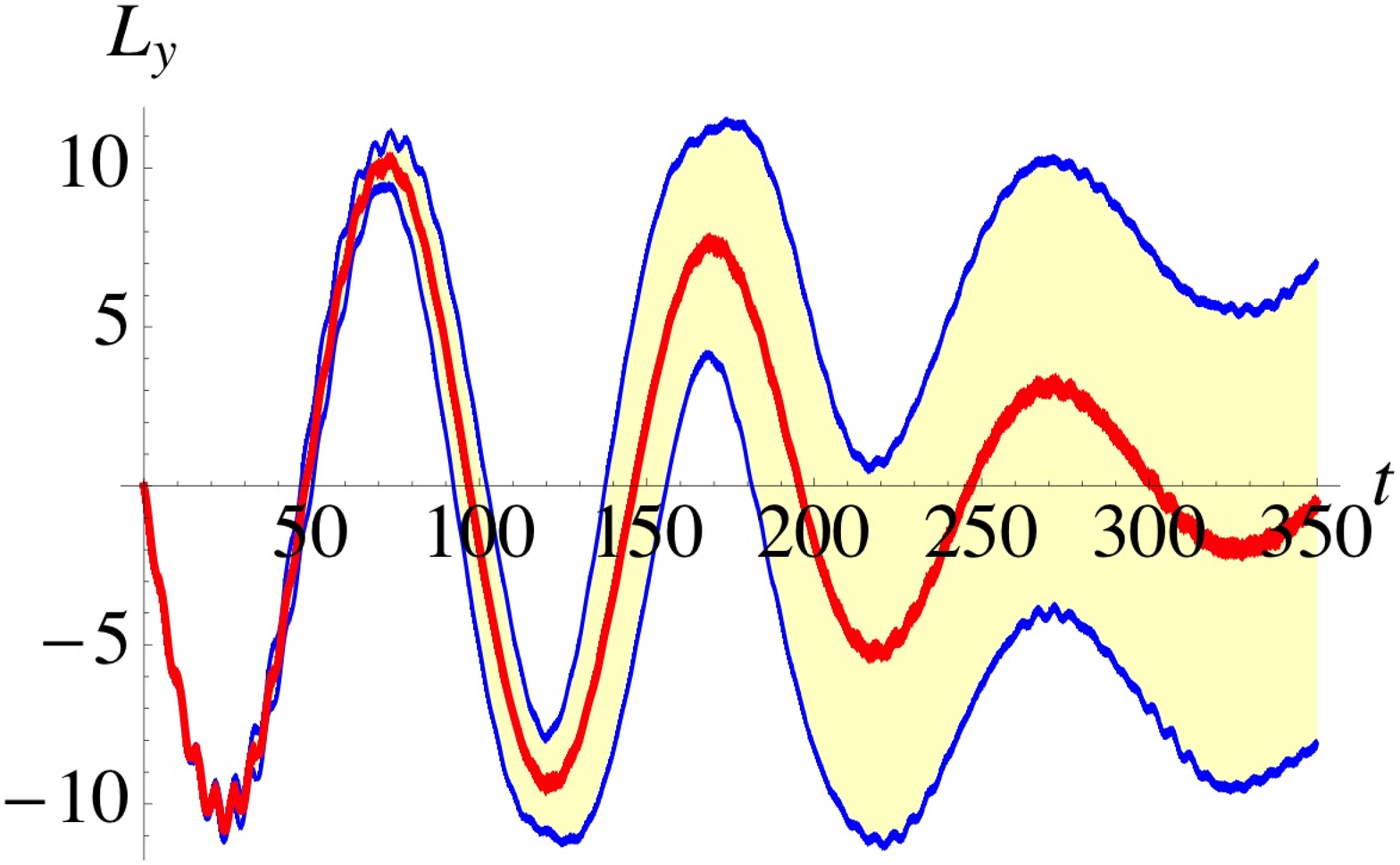}}
\caption{(Color online) Stochastic dynamics with $\varepsilon_x(t)$,
$\varepsilon_y(t)$ and $\varepsilon_z(t)$ fields taken as Gaussian
white noise.  There is little change from the results shown in
Figs.~\ref{Fig_n_vs_t_in_E_field_m_v} and
\ref{Fig_L_vs_t_in_E_field_m_v}.}
\label{Fig_vareps_z_m_v}
\end{figure}

In Fig.~\ref{Fig_vareps_z_m_v} we also allowed the $z$-component of
the electric field to fluctuate, i.e., we allowed $\varepsilon_z(t)$
to be a non vanishing stochastic variable with ``volatility''
(standard deviation) $\varepsilon_0 = 0.1$.  Clearly, there is not
very much of a change due to $\varepsilon_z(t)$.  Again, despite the
fluctuations of the electric field, $\overline{n_z(t)}$ and
$\overline{L_z(t)}$ do not decay to zero as do the other components of
$\overline{{\bf n}(t)}$ and $\overline{{\bf L}(t)}$.  This is in
contrast to the motion of spin in a stochastic magnetic field, where
all the spin components decay to zero for Gaussian white noise in all
the magnetic field components \cite{STB_2013}.


\section{Summary and Conclusion}   \label{Sec:Summary}

We introduced a model for treating the dynamics of an electric dipole
moment in the presence of a deterministic electric field and an
environment with which the dipole interacts.  Environmental
decoherence was modeled by considering a stochastic fluctuating
electric field (noise) which interacts with the electric dipole
moment.  We solved the stochastic mean-field equations of motion for
Gaussian white noise.  The model makes the {\it external noise}
assumption \cite{vanKampenBook} wherein no back-action of the system
on the environment is present.  A consequence of this assumption is
that the system does not come into equilibrium with a thermal
environment, but goes to the most democratic density matrix state
having zero expectation value of the dipole moment \cite{STB_2013}.
This is a good approximation when the back-action is weak, as
explained in \cite{vanKampenBook, STB_2013}.  But even if it is not
weak, one way of overcoming this problem is to augment the equations
of motion for the electric dipole moment with a decay term that
insures that the system comes into thermal equilibrium at long times.
If we schematically represent the equation of motion for the dipole
moment as, $\frac{d{\bf d}}{dt} = {\cal O} {\bf d}$, and add a decay
term $\eta$ to get the augmented equation of motion, $\frac{d{\bf
d}}{dt} = {\cal O} {\bf d} - \eta$, then at large times, we can set
the rate of change of the dipole moment to be zero and the dipole
moment to its equilibrium value as given by a Boltzmann averaged
dipole moment, ${\bf d}_{\mathrm{eq}} = {\mathrm{Tr}} [e^{-\beta
H}{\bf d}]/{\mathrm{Tr}} [e^{-\beta H}]$, where $\beta$ is the inverse
temperature of the bath.  Hence, as $t \to \infty$, we find that $\eta
= {\cal O} {\bf d}_{\mathrm{eq}}$.  Thus, the augmented equation of
motion becomes,
\begin{equation} \label{dot_p_aug}
    \frac{d{\bf d}}{dt} = {\cal O}(t) {\bf d}(t) - {\cal O} {\bf
    d}_{\mathrm{eq}} .
\end{equation}
This equation yields the right thermal equilibrium result
asymptotically, ${\bf d}(t) \xrightarrow[{t \to \infty }]{} {\bf
d}_{\mathrm{eq}}$.  Similarly for the angular momentum equation,
\begin{equation} \label{dot_L_aug}
    \frac{d{\bf L}}{dt} = \tilde {\cal O}(t) {\bf d}(t) - \tilde {\cal
    O} {\bf L}_{\mathrm{eq}} ,
\end{equation}
where ${\bf L}_{\mathrm{eq}} = {\mathrm{Tr}} [e^{-\beta H}{\bf
L}]/{\mathrm{Tr}} [e^{-\beta H}]$.  This approach may be overly
simplistic if multiple decoherence processes play a role in the
back-action dynamics, but it does yield dynamics that tend
asymptotically to the correct equilibrium results when 
back-action is not negligible.

Here, we showed that the dynamics of an electric dipole moment in a
stochastic field is more complicated than the dynamics of a magnetic
dipole moment in a stochastic magnetic field.  Even with the external
noise assumption, and even for Gaussian white noise, not all the
components of the average electric dipole moment and the average
angular momentum decay to zero, despite fluctuations in all three
components of the electric field.  This is in contrast to the decay of
the average over fluctuations of a magnetic moment, which does decay
to zero in a stochastic magnetic field with Gaussian white noise in
all three components \cite{STB_2013}.  Here, $\overline{L_z(t)}
\xrightarrow[{t \to \infty }]{} L_z(0)$, and $\overline{n_z(t)}$,
which is proportional to the Stark energy, also does not decay to zero
at large times; the system does not come into equilibrium.  These
predictions, which are valid under the external noise assumption,
should be able to be readily checked experimentally.  The predictions
will remain valid also for Gaussian colored noise stochastic process,
as long as the temporal correlation time of the noise process,
$\tau_c$, is short compared with the rotation time of the molecule,
$\tau_r = I/\overline{L}$, and the Stark timescale, $\tau_S = \hbar/(E
d)$.

\begin{acknowledgments}
This work was supported in part by grants from the Israel Science
Foundation (No.~2011295) and the James Franck German-Israel Binational
Program.  Useful discussions with Yshai Avishai and Yehuda Ben-Shimol
are gratefully acknowledged.
\end{acknowledgments}


\begin{thebibliography}{99}
\bibitem{sym_top}
For simplicity we take a symmetric top, rather than a spherical top or
an asymmetric top Hamiltonian.
\end{thebibliography}
\end{document}